# Assessment of Hybrid RANS-LES and WALE Formulations for Wake and Resistance Prediction of the BB2 Submarine

**Noh Zainal Abidin[*], Frederic Grondin[*], Pol Muller[†] and Jean-François Sigrist[‡]**
[*]GEM ECN, Nantes/France, [†]Sirehna-Naval Group, France, [‡] eye-PI, France
noh.bin-zainal-abidin@ec-nantes.fr, frederic.grondin@ec-nantes.fr

## 1 Introduction

Submarine hydrodynamics presents unique challenges in accurately predicting flow separation, wake structure, and resistance due to complex geometry and turbulent behaviour at high Reynolds (Re) numbers. Traditional Reynolds-Averaged Navier-Stokes (RANS) approaches are often limited in resolving unsteady flow structures and turbulence in the near and far regions. To address these limitations, hybrid RANS-LES models such as Detached Eddy Simulation (DES) and Large Eddy Simulation (LES) offer improved performance in capturing near-wall vortical structures. The capturing of turbulent vortices and wake structures significantly contributes to conduct hydrodynamic noise analysis. Detailed resolution and understanding of these coherent structures help minimize hydroacoustic signatures, essential for submarines' stealth characteristics. Based on prior studies, Breuer et al. (2003) reported that RANS failed to capture unsteady vortex shedding, producing only steady results even in 3D simulations. In contrast, DES and LES successfully resolved asymmetric shedding across different grid resolutions. Spalart (2009) reported that DES is more effective than RANS or LES for high Re flows, although it suffers from challenges related to ambiguous grids and nonmonotonic grid refinement behaviour. Whereas Liang & Xue (2014) found that DES predicts tip vortex flow characteristics more accurately than RANS-SA and can capture complex 3D vortex structures. In addition, Guilmineau et al. (2018) demonstrated that the IDDES model accurately predicts recirculation bubbles and aligns more closely with experimental data for flow prediction. Long et al. (2021) confirmed the capability of DDES in simulating cavitating flows around hydrofoils and marine propellers. Lungu (2022) highlighted the efficiency and accuracy of the hybrid IDDES-SST model in DARPA submarine simulations. Zhang et al. (2023) also noted that URANS struggles with resolving small-scale turbulence structures, whereas IDDES is better suited for predicting complex phenomena such as ship air wake asymmetry. Nevertheless, capturing the unsteadiness and turbulence fluctuations scales is extremely challenging because the cell size requirements should suit each turbulence model employed. Thus, as a continuation of the prior work of Abidin et al. (2024), the unsteady simulation with high mesh resolution at $U_m$=1.8235 m/s and Re of 3.6×10$^6$ to generate the asymmetrical wake dynamic and vortical structure. The current research expands the methodology by parameterizing the meshes and numerical scheme based on Taylor microscale refinement with respect to the characteristic length of ($L$, $B$ and $D$) of submarine, particularly focusing on hybrid turbulence model and WALE to observe the ability to resolve turbulence in the wake region. The transient simulations were performed initially using wall-resolved mesh (76×10$^6$ cells) at $y^+$ < 5 and then wall-modelled mesh (56×10$^6$ to 74×10$^6$ cells) at $y^+$ > 30, which produced notably different and more detailed results (vortices and turbulence fluctuation) than previous steady-state RANS simulations without risking the accuracy of quantity of interest (global resistance).

## 2 Numerical Test Case

The scaled BB2 submarine is utilized in the present work based on the availability of experimental and numerical research databases. For instance, the particle image velocimetry measurements were performed by Kumar et al.(2012), while the hydrodynamic forces measurements on the hull were reported by Quick & Woodyatt (2014) and Fureby (2017). Besides that, the NATO AVT-301 collaboration group also conducted various CFD studies on the BB2 submarine. The BB2 submarine designed by the Maritime Research Institute Netherlands (MARIN) in a variant of the Joubert (2006), has a $L/D$ ratio of 7.3. Based on Bettle (2014) and Overpelt et al. (2015), the modifications enhance submarine stability and control. Details of the scaled BB2 submarine can be found in Abidin et al.

(2024). MARIN provided the 3D CAD of the full-scale submarine, while Sirehna-Naval Group supplied lab-scale data for validation.

## 3 Mesh and Numerical Scheme Parameterization

The mesh convergence studies can be found in Abidin et al. (2024). Now, the project study focuses on parameterizing the mesh and numerical scheme to readily capture the turbulence fluctuation scale efficiently. The mesh parameterization of refinement zone cell size at sail, upper mid body and rudder zone of submarine as shown in Fig.1, based on the Taylor microscale, $\lambda_T = \sqrt{15} \cdot A^{-1/2} \cdot R_l^{-1/2} \cdot l$, (Tennekes & Lumley,1994) and explained by Howard & Pourquie (2002) and Guilmineau et al. (2018) is utilized concerning characteristic length ($L$ = 2 m, $D$ = 0.4615 m, $B$ = 0.2737 m). Where $A$ is non-dimensional constant, 0.5 while $R_l$ is the Reynold number corresponding to characteristic length chosen. The Taylor microscale is intermediate length scale at which fluid viscosity significantly affects the dynamics of turbulent eddies in the flow and appropriate for DES and LES simulations. The size of computational domain utilized was $1L$ (front) x $1L$ (lateral) x $3L$ (wake) as explained in Abidin et al. (2025). Here we are targeting the mesh resolution in range $50 \times 10^6$ to $76 \times 10^6$ cells for straight ahead course to have affordable computational load in HPC without sacrificing the vortex shedding and wake dynamic. Thus, by employing Cadence Fidelity tool able to control smoothness which gradually expands the size of hexahedral mesh effectively. Thus, only two refinement regions are made near to the wall with ratio of $RZ1$ of depth, $d1/L$= 0.325, breadth, $b1/L$ =0.2 and length, $l1/L$ = 1.05. While for $RZ2$ (imposed Taylor Microscale), $d2/L$= 0.2775, $b2/L$ =0.15 and, $l2/L$ = 1. Several test cases on mesh parameterization as shown in Table.1.

Table 1. The cell size for $RZ2$ (Taylor Microscale)

| Case | $l$ (m) | $R_l$ | $\lambda_T$ (mm) |
|---|---|---|---|
| 1 | 2 | $3.6 \times 10^6$ | 5.76 |
| 2 | 0.4615 | $0.8 \times 10^6$ | 2.77 |
| 3 | 0.2737 | $0.5 \times 10^6$ | 2.13 |

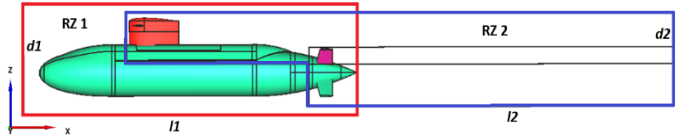

Fig.1. The refinement zone placement from lateral view ($RZ1$ and $RZ2$).

Initially, the mesh generated in wall-resolved and wall-modelled with Case 1, $\lambda_T$=5.76 mm and Case 2, $\lambda_T$=2.77 mm to inspect the generation of turbulence fluctuation and asymmetric in the wake during simulation. All these cases will generate the high mesh resolution from $50 \times 10^6$ cells to $76 \times 10^6$ cells. The CFD application OpenFOAM 11 (OpenFOAM, 2023) was used in the current study. The simulation uses a second-order implicit backward scheme for time discretization. To ensure capturing the small scale of turbulence fluctuations, the maximum CFL = $U \cdot \Delta t / \Delta x$ of less than 2 implemented as suggested by Rocca et al. (2022). Hence, the $\Delta t$ =1×10$^{-7}$ s for wall-resolved mesh and $\Delta t$ =1×10$^{-6}$ s for wall-modeled mesh utilized. Statistics were computed from data sampled over the submarine for a duration of $t_c$ = 10 where the $t_c = t.U/L$ requiring significantly large HPC computing time and resources with capacity of 500 cores/qos at 4 Gb/core in GLiCID Computing Facility, Nautilus for four to five months computation. The Gradients terms are computed with a second-order cellLimited Gauss linear scheme. Divergence terms use second-order Linear-Upwind Stabilized Transport (LUST) for $\nabla \cdot (\phi U)$ and Limited linear for $\nabla \cdot (\phi k)$ and $\nabla \cdot (\phi \tilde{v})$, offering second-order accuracy in smooth regions. Viscous terms apply a second-order Gauss linear scheme. Laplacian and surface-normal gradient terms use Gauss linear limited corrected 0.5, also second-order. Linear interpolation is applied, and wall distance is evaluated using the Mesh-wave method. Pressure is solved using PCG with DIC preconditioner, tolerance 1×10$^{-6}$, and relative tolerance 0.01. The final pressure solution uses zero relative tolerance. Velocity, turbulence, and related fields used PBiCGStab with DILU, tolerance 1×10$^{-8}$, and relative tolerance 0.001. The PIMPLE loop runs for 3 outer loop of the PIMPLE algorithm, 3 inner corrector loops for pressure-velocity coupling, and 1 non-orthogonal corrector. Full relaxation (factor = 1) is applied to $U$ and $\tilde{v}$. Scale-resolving simulations (SRS) were performed for straight-ahead and initialized by RANS solution (prior study) and extended using hybrid turbulence models of DDES

Spalart-Allmaras and IDDES Spalart-Allmaras, and WMLES (WALE). Due to the modelled stress depletion (MSD) effect observed in the DES turbulence model, which reduces its accuracy, the results from DES are not presented in this work (Spalart, 2009). The details of turbulence models implemented can be found within the OpenFOAM source code at *OpenFOAM-11/src/MomentumTransportModels*.

## 4 Results and Discussion

Initially, two mesh configurations (Fig. 2 and Fig. 3) were evaluated to examine the influence of wall resolution on asymmetric wake formation. However, both Case 1 ($\lambda_{T1}$=5.76 mm) and Case 2 ($\lambda_{T2}$=2.77 mm), which were scaled based on reference lengths $D$ or $L$, failed to capture adequate turbulence fluctuations. Despite being wall-resolved, these meshes resulted in quasi-steady flow patterns and underdeveloped wake structures, as shown in Fig. 4(a). Only limited unsteadiness appeared near the rudder tip vortex, and the flow behaviour did not align with the experimental observations reported by Chen et al. (2023). In contrast, Case 3, where mesh refinement was based on the width ($B$), demonstrated sufficient resolution to reproduce the wake dynamics with higher fidelity. Interestingly, the wall-resolved mesh and the wall-modelled mesh produced comparable wake structures. The primary difference lies in the computational cost, whereas enforcing CFL < 2 for the wall-resolved mesh required a prohibitively small time step of $\Delta t$ =1×10$^{-7}$ s. In contrast, the wall-modelled mesh allowed a more practical time step of $\Delta t$ =5×10$^{-6}$ s. Consequently, for further assessment using hybrid turbulence models and the WALE model, the wall-modelled mesh configuration with approximately 56×10$^{6}$ and 74×10$^{6}$ cells was adopted due to its balance between accuracy and computational efficiency.

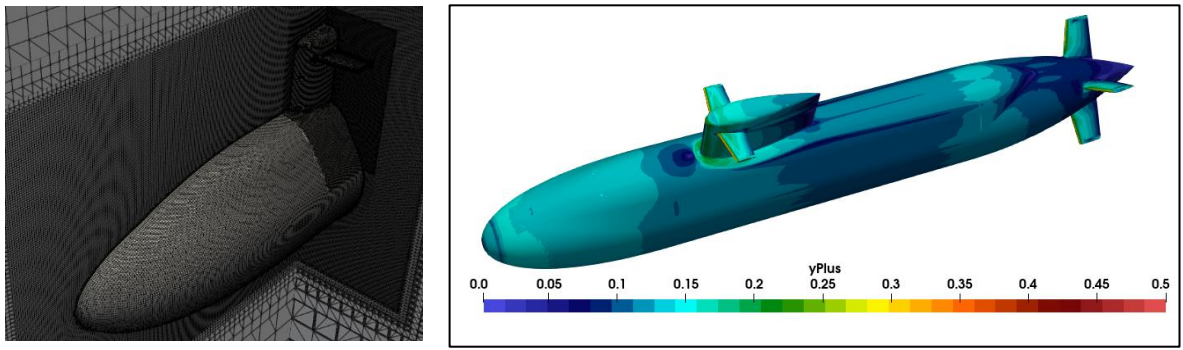

Fig.2. Wall resolved Mesh $y^+$ < 5 (76×10$^6$ cells) - Case 1

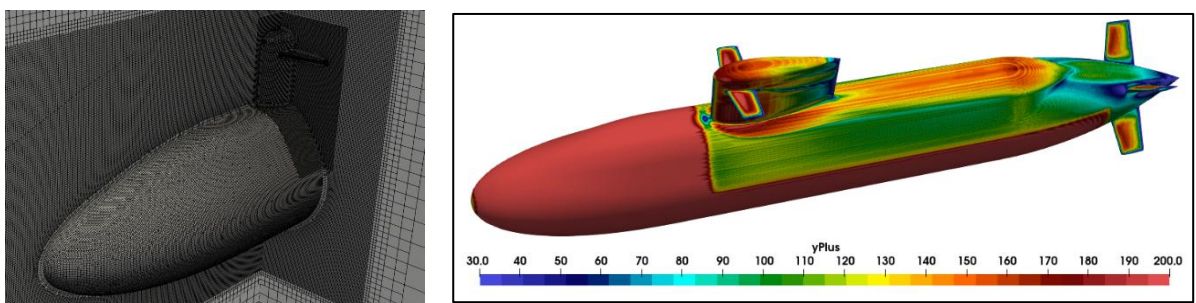

Fig.3. Wall-Modelled Mesh $y^+$ > 30 (56×10$^6$ cells) - Case 1

Figures 4(b–d) show asymmetric wake dynamics using DDES, IDDES, and WALE models with a wall-modeled mesh (Case 3). Figure 4(b), using 56×10$^6$ cells, the mesh refinement covers up to the rudder and does not extend to $l2$; wake structures from sail, hydroplane, and rudder are clearer than in Fig. 4(a), but still limited by coarse wake resolution. Figures 4(c–d) use a refined 74×10$^6$ cell mesh extended to $l2$, enabling better RANS–LES switching in IDDES and more accurate capture of flow separation, transition, and vortex dynamics. The WALE model in Fig. 4(d) with 74×10$^6$ cells illustrated the detail in resolving small-scale turbulent structures, particularly in the wake and near the sail region. The flow appears more chaotic and fully developed, demonstrating WALE's strength in capturing vortex

breakdown and separation in the LES region without excessive near-wall damping. Nevertheless, despite using the same mesh resolution as IDDES, there is a deficiency in the wake region near the bottom of the rudder, which exhibits a shorter and less energetic wake. Potentially, the WALE model's eddy-viscosity formulation smooths out flow features more aggressively in that specific zone, leading to early dissipation of vortices or insufficient triggering of instabilities near the rudder bottom zone. In contrast, the IDDES model generates a more developed wake structure in this area and shows extended and energetic vortex structures originating from the rudder base, suggesting active flow separation and strong interaction between the rudder and the hull wake. The model improved RANS-LES blending, which allows it to resolve this transitional region better and sustain coherent vortical structures downstream. The vortex structures identified in Fig. 4 are further examined in Figs. 5 and 6 through the normalized mean axial velocity, $\overline{u_x}/U$ and normalized mean vorticity magnitude, $\overline{\omega}.r/U$ distribution on various cross-sections, from mid-body to the stern ($x/L$ = 0.475 to 1) as referring to Chen et al. (2023) and Visonneau et al., (2020) procedure. The top row (a) shows the RANS solution with a wall-modelled mesh, serving as a reference. In Fig.5, the flow across all models of DDES, IDDES, and WALE is symmetric up to $x/L$ = 0.75. Beyond this, IDDES presented the widest wake and strongest velocity deficit, indicating enhanced turbulent mixing. DDES captures a narrower, weaker wake, while WALE shows more symmetry but reduced deficit, suggesting earlier dissipation and lower turbulence intensity. While in Fig.6 highlighted the ability of each turbulence models to capture rotational flow and shear-layer dynamics. It can be seen that all models presented minimal vorticity up to $x/L$ = 0.75. Beyond this, IDDES (c) captures the most intense and widespread vorticity at $x/L$ = 0.95 and 1.0, indicating strong vortex shedding and turbulent mixing. DDES (b) reveals weaker and more localized vorticity, suggesting under-resolved wake structures especially on capturing the hydroplane tips vortex (HTV). While WALE (d) maintains clearer symmetry, exhibits reduced intensity at horseshoe vortex (HSV) and improved capturing HTV and rudder tips vortex (RTV) compared to IDDES and DDES. This confirms that WALE and IDDES best capture near-wake turbulence and vortex dynamics, while DDES and produce more dissipative or underdeveloped wakes. Fig.7 illustrated the time evolution of viscous ($F_v$), pressure ($F_p$), and total resistance ($F_t$) forces from the IDDES simulation. After initial transients, all forces stabilize around $t_c$=10, indicating convergence achieved. The $F_t$ aligning well with experimental data, with slight overprediction driven mainly by pressure forces. In addition, Table 2 compares global resistance predictions across turbulence models. RANS k-ω SST and WALE yield the lowest errors (0.3% and 1.1%), while IDDES and DDES slightly overpredict due to stronger wake and pressure drag effects. Overall, IDDES provides a good balance between accuracy and wake resolution fidelity.

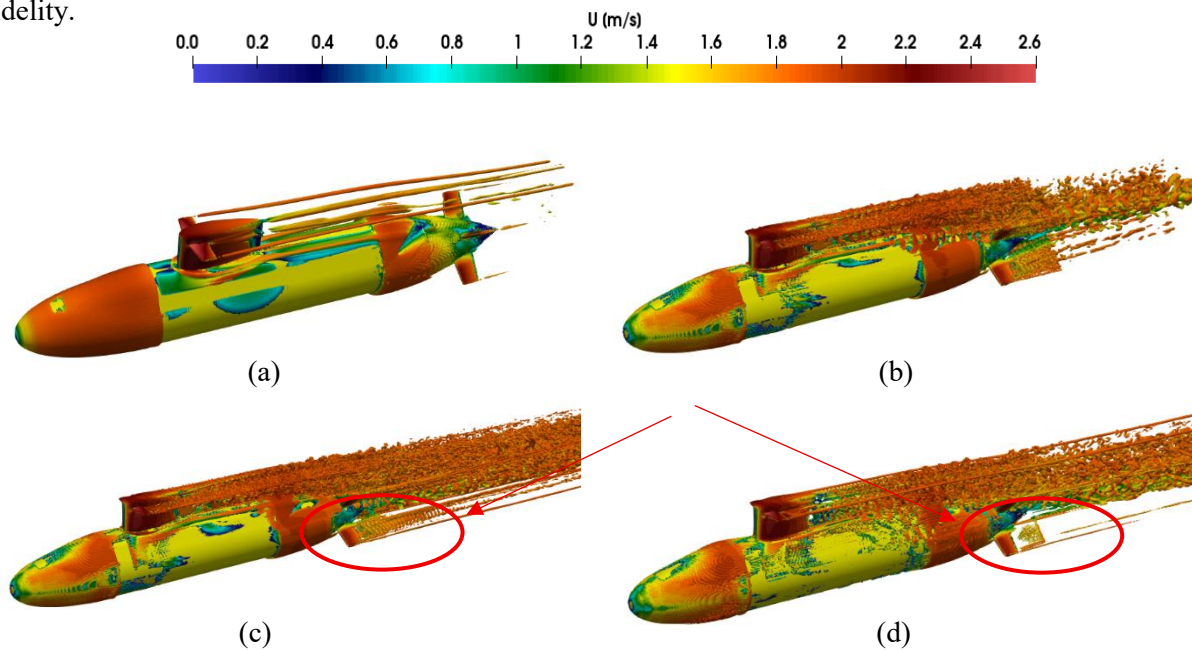

Fig.4. Q-criterion iso-surfaces colored by instantaneous velocity, (a) Case 1 and 2 (DDES), (b) DDES (56×10⁶ cells), (c) IDDES (74×10⁶ cells), (d) WALE (74×10⁶ cells)

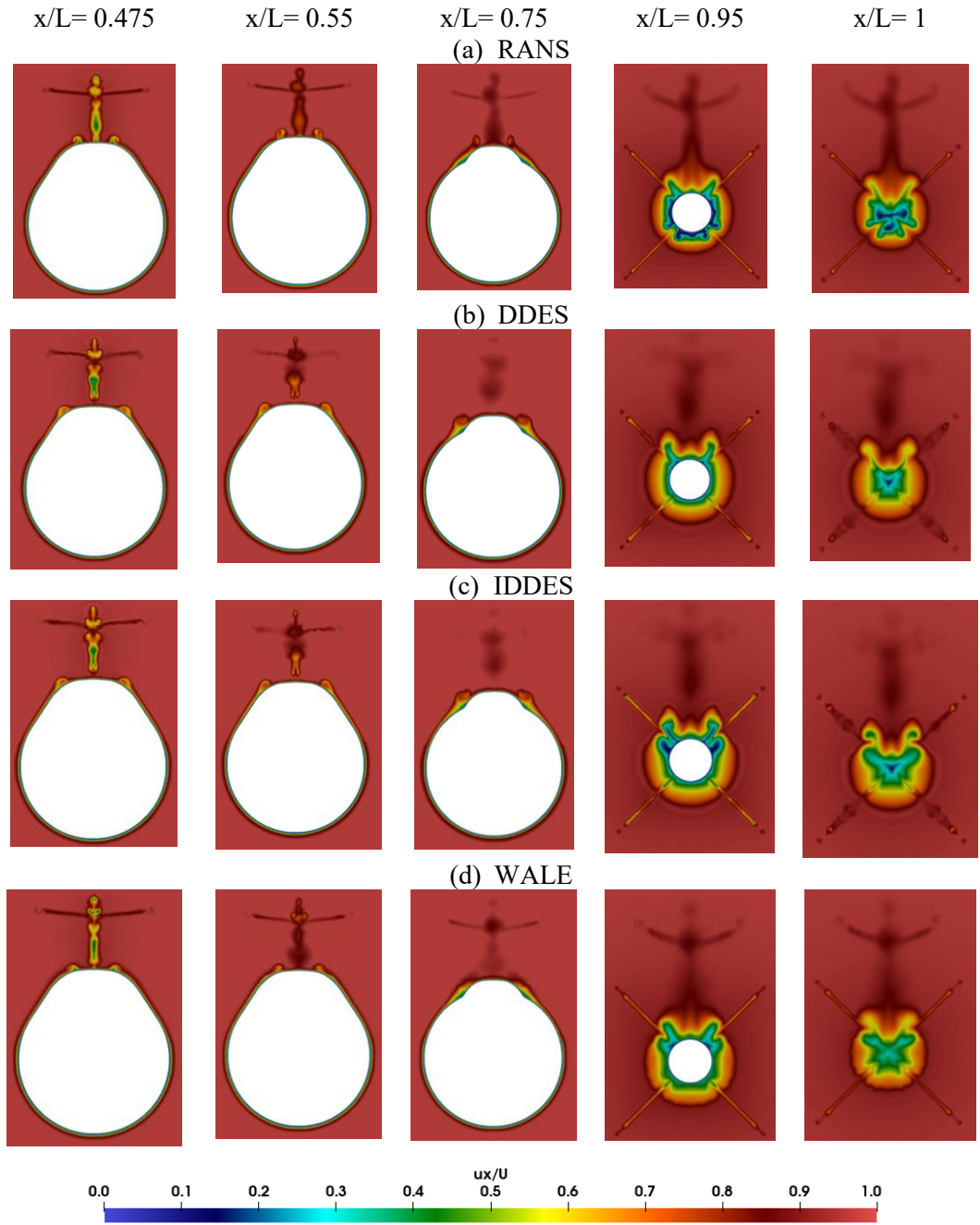

Fig.5. Normalized Mean Axial Velocity at each section of submarine from nose (x/L=0) to rudder (x/L=1), (a) RANS (56×10$^6$ cells), (b) DDES (56×10$^6$ cells), (c) IDDES (74×10$^6$ cells), (d) WALE (74×10$^6$ cells)

## 5 Conclusion

RANS SST and WALE models yield the most accurate global resistance values, closely matching experimental data. However, RANS underpredict wake turbulence, producing a steady state solution, and WALE has limitations on capturing the bottom rudder wake. DDES improves unsteady flow capture but overpredicts resistance due to limited wake resolution. IDDES offers the best overall performance, balancing resistance accuracy with detailed wake structure prediction, making it the most suitable for submarine hydrodynamic analysis. Future research should investigate energy spectrum development and acoustic-related phenomena in the wake, particularly behind the rudder, to better understand unsteady flow mechanisms in complex submarine geometries.

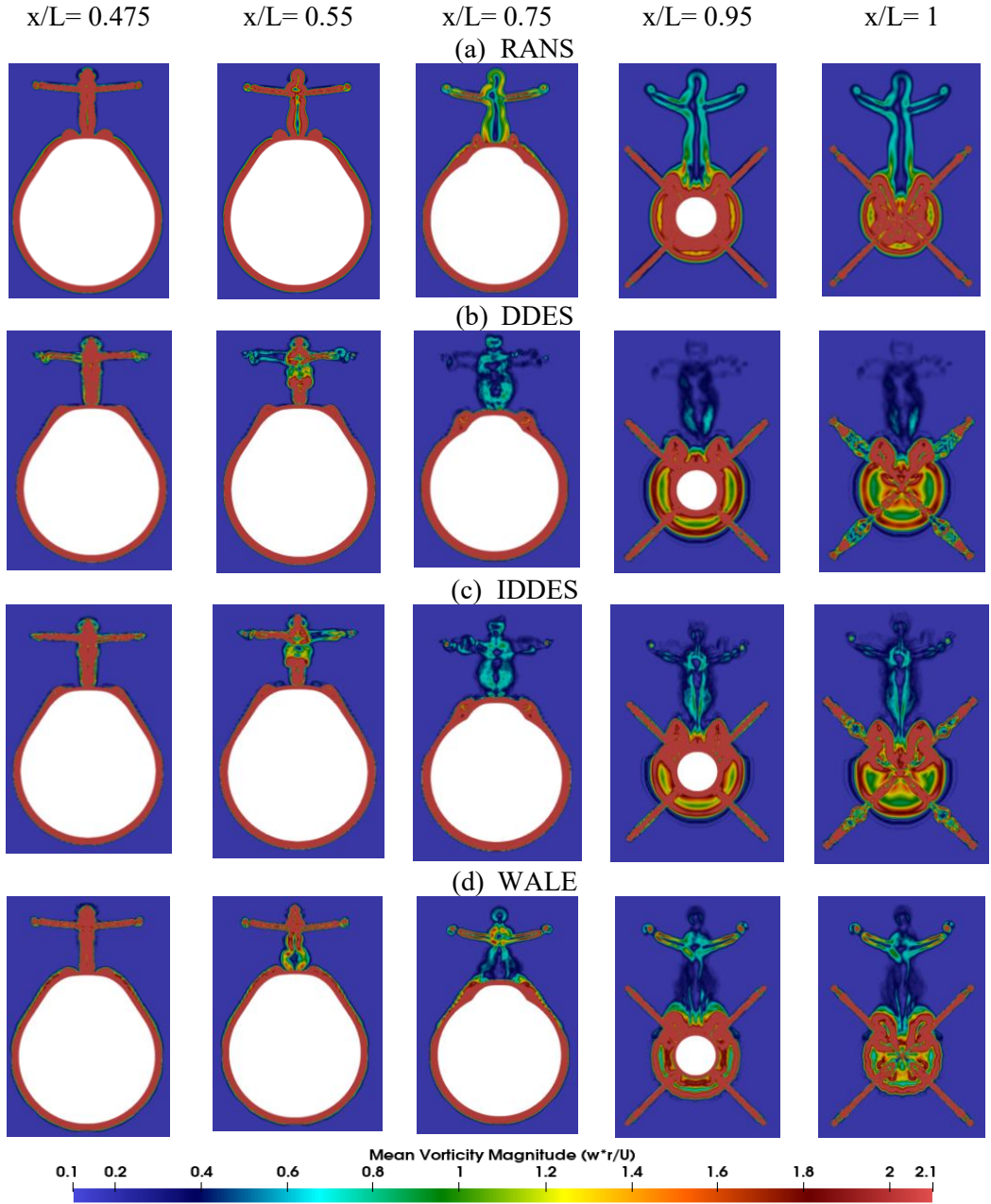

Fig.6. Normalized Mean Vorticity Magnitude at each section of submarine from nose (x/L=0) to rudder (x/L=1), (a) RANS (56×10$^6$ cells), (b) DDES (56×10$^6$ cells), (c) IDDES (74×10$^6$ cells), (d) WALE (74×10$^6$ cells)

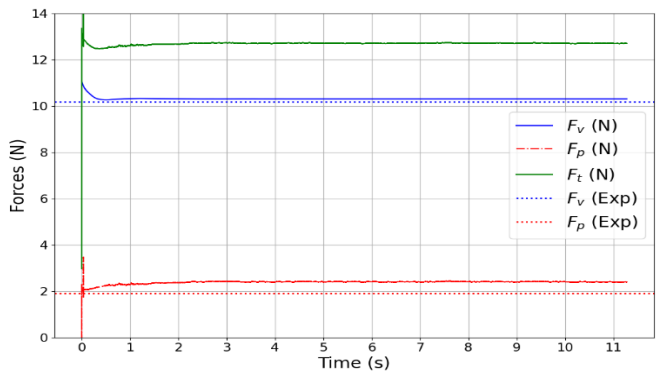

Fig.7. Local and Global Resistance (IDDES) at $t_c$ = 10

Table 2. Comparison of Global Resistance of turbulence models

| Turbulence Model | $F_t$ (N) | $e_{exp}$(%) |
|---|---|---|
| RANS $k$-$\omega$ SST | 12.03 | 0.6 |
| DDES | 12.65 | 4.6 |
| IDDES | 12.59 | 4.1 |
| WALE | 12.23 | 1.1 |

**Acknowledgements**

This work is funded under the Offset Program LCS between Naval Group, ECN, and NDUM. The authors thank Sirehna-Naval Group for technical support and access to databases, Dr. Joel Guerrero (Wolf Dynamics Srl) for assistance with OpenFOAM, and Dr. Emmanuel Guilmineau (ECN-LHEEA) for support in post-processing submarine hydrodynamic features. This research used HPC resources of the GLiCID Computing Facility (Ligerien Group for Intensive Distributed Computing, https://doi.org/10.60487/glicid, Pays de la Loire, France).